\documentclass[pdflatex,sn-mathphys-ay]{sn-jnl}

\usepackage{graphicx}
\usepackage{multirow}
\usepackage{amsmath,amssymb,amsfonts}
\usepackage{amsthm}
\usepackage{mathrsfs}
\usepackage[title]{appendix}
\usepackage{xcolor}
\usepackage{textcomp}
\usepackage{manyfoot}
\usepackage{booktabs}
\usepackage{algorithm}
\usepackage{algorithmicx}
\usepackage{algpseudocode}
\usepackage{listings}

\usepackage{fontawesome5}  
\usepackage{hyperref}      

\newcommand{\orcidicon}[1]{%
  \href{https://orcid.org/#1}{\textcolor[HTML]{A6CE39}{\faOrcid}}%
}

\usepackage{xcolor}
 
\newcommand{\fracbrac}[2]{\left(\frac{#1}{#2}\right)}

\theoremstyle{thmstyleone}%
\theoremstyle{thmstyletwo}%

\theoremstyle{thmstylethree}%

\begin{document}

\title[Closed Form Expressions for the Potentials and Accelerations of Generalized Ring Models]{Closed Form Expressions for the Potentials and Accelerations of Generalized Ring Models}

\author*[1]{\fnm{Zachary} \sur{Murray}\orcidicon{0000-0002-8076-3854}}\email{zachary.murray@geoazur.unice.fr}

\affil*[1]{\orgdiv{Geoazur}, \orgname{Université Côte d'Azur}, \orgaddress{\street{Av. Valrose}, \city{Nice}, \postcode{06000}, \country{France}}}

\abstract{We present several closed form expressions of useful mass distributions. These include the potentials and accelerations of circular rings and arcs, the potentials of uniform density rings and arcs at arbitrary eccentricities, and the potentials and accelerations of rings and arcs when the mass is time-averaged over a Kepler orbit.  We show that these expressions can be expressed, often simply, in terms of elliptic functions of complex arguments.  We show that in a few limiting cases, the expressions are entirely real.  We expect that these expressions will allow for more rapid modeling in many areas of celestial mechanics.}


\maketitle

\section{Introduction}\label{sec1}
Hierarchal gravitational systems are common in the universe, examples include protoplanetary disks, exoplanetary systems, the solar system as a whole, lunar systems and others.  Consequently, objects following nearly Keplerian orbits or those lying at the equilibria points produced by such orbits (e.g. the Lagrange points) are common. To study the long-term dynamics and model systems of small bodies, it is necessary to derive expressions for their gravitational fields that, when averaged over time, often correspond to mass distributions along circular and eccentric orbits. Ideally, both the potentials and the accelerations due to such mass distributions would be expressed in closed form, since the former is useful for analytic work and the latter for numerical simulations. 

A few such expressions already exist.  For example, the potential of a simple circular ring \citep{Lass_1983,Fukushima_2010}
and its associated dynamics \citep{Broucke_2005,Schumayer_2019,Igata_2020} are well studied and have been used to model the Main Asteroid Belt \citep{Kuchynka_2010,Gomes_2023} and Kuiper Belt \citep{Pitjeva_2018,Park_2021}.  In more detailed studies, several inclined rings have also been used \citep{Liu_2022}.

However, in general, expressions for the potentials and accelerations of more complex distributions of mass, including arcs and eccentric rings are still relatively underdeveloped.  While multipole or other series expansions can are often employed, they are typically only valid at moderate eccentricities.  These models often fail to describe the dynamics at higher eccentricities, when the dynamics leaves the regime of validity of the approximation.  Typically at higher eccentricities models reveal new phenomena or break down \citep[e.g.,][among others]{Li_2014,Naoz_2017,Wang_2017,Murray_2022}.

In this paper, we derive closed-form expressions in terms of elliptic integrals for the gravitational potentials of circular arcs, uniform eccentric rings, and eccentric rings/arcs that mimic the mass distribution of averaged Kepler orbits.  We also provide accelerations for the circular and keplerian cases.  We present a variety of numerical checks to verify our work and show agreement with the literature for limiting cases.   Given previous work on optimizing the computation of elliptic integrals \citep{Fukushima_1994,Fukushima_2008,Fukushima_2009,Fukushima_2010,Fukushima_2011,Fukushima_2012}, these expressions allow both for rapid computation and analytic insight.


\section{Uniform Circular Arcs}

The basic problem in this work is to compute the acceleration at some arbitrary point in space, due to a geometric object, and to provide that form in an quickly computable format. It is most convenient to do this by describing the gravitational potential per unit mass $V$ of the object via 
\begin{equation}
    V = - G \int_M \frac{1}{|\vec{x}-\vec{r}|} dM
\end{equation}
where $M$ describes the mass distribution in question. The corresponding accelerations on an arbitrary point due to these distributions, given by the gradient of the potential 
$\vec{a} = \nabla \phi$

Hence, the principal difficulty in computing a potential or acceleration is finding a way to rapidly compute a difficult integral.  To begin, we wish to solve for the potential everywhere, given an arc-shaped mass distribution.  We assume that the arc lies in the x-y plane and assume a cylindrical coordinate system.  We describe the arc as having a radius $a$, a central theta $\Theta_0$, and an angular span $\delta \theta$, such that $\delta \theta = \pi$ corresponds to the circular limit.  The arc is assumed to have mass $M$ (and hence linear mass density $\lambda = M/2 \delta \theta r$).  Finally, we evaluate the potential at some arbitrary radial coordinate $r_0$, theta coordinate $\theta_0$ and z coordinate $z_0$. The integral of the potential becomes:

\begin{equation}
V = - G \lambda \int_S \frac{r d\theta}{|\vec{x}-\vec{r}|} = -\frac{G M}{2\delta \theta } \int_{\Theta_0 - \delta \theta}^{\Theta_0 + \delta \theta} \frac{1}{\sqrt{a^2 + r_0^2 + z_0^2 - 2 a r_0 \cos(\theta - \theta_0)}} \, d\theta
\label{eq:arcbase}
\end{equation}

This integral has a closed form in terms of elliptic integrals.   The form becomes particularly simple when given the following definitions:

\begin{subequations}\label{eq:defarc}
\begin{align}
R &= \sqrt{r_0^2 + z_0^2}, \\
C_1 &= \sqrt{a^2 + R^2 - 2 a r_0 \cos(\delta\theta + \theta_0 - \Theta_0)}, \\
C_2 &= \sqrt{a^2 + R^2 - 2 a r_0 \cos(\delta\theta - \theta_0 + \Theta_0)}, \\
d &= \sqrt{a^2 - 2 a r_0 + R^2}, \\
p &= \sqrt{a^2 + 2 a r_0 + R^2}, \\
q^2 &= a^2 - 2 r_0^2 + R^2, \\
k &= -\frac{4 a r_0}{a^2 - 2 a r_0 + R^2}.
\end{align}
\end{subequations}

Note here we define $q^2$ rather than $q$ itself to avoid imaginary values for when $a^2 + R^2 < 2 r^2$. Since only the $q^2$ term appears in our expressions for the potential and accelerations, this means that both can be computed without resorting to imaginary quantities.  When expressed in terms of these variables, the potential can be written as 

\begin{equation}
V = \frac{G M}{\delta\theta \, d} \left( \textbf{F} \left( \frac{-d\theta - \theta_0 + \Theta_0}{2} , k \right) - \textbf{F} \left( \frac{(d\theta - \theta_0 + \Theta_0)}{2} , k \right) \right).
\label{eq:arcpot}
\end{equation}

Where $\textbf{F}(x,k)$ is the incomplete elliptic integral of the first kind.  Now that we have a closed form for the potential is found, the corresponding accelerations can be derived by simply taking the gradient.  In cylindrical coordinates these alignments become (using the definitions in Eq \ref{eq:defarc})

\begin{subequations}\label{eq:arc_acc}
\begin{align}
a_z &= -\frac{G M z}{d p^2 \delta\theta } \Bigg( 
\textbf{E} \left( \tfrac{1}{2} (\delta\theta + \theta_0 - \Theta_0), k \right) + 
\textbf{E} \left( \tfrac{1}{2} (\delta\theta - \theta_0 + \Theta_0), k \right) \notag \\
&\qquad +
\frac{2 a r}{C_1 d} \sin(\delta\theta + \theta_0 - \Theta_0) + 
\frac{2 a r}{C_2 d} \sin(\delta\theta - \theta_0 + \Theta_0)
\Bigg) \\
a_r &= \frac{G M}{2 d^2 \delta\theta } \Bigg( 
\frac{d q^2}{p^2 r} \textbf{E} \left( \tfrac{1}{2} (\delta\theta + \theta_0 - \Theta_0), k \right) + 
\frac{d q^2}{p^2 r} \textbf{E} \left( \tfrac{1}{2} (\delta\theta - \theta_0 + \Theta_0), k \right) \notag \\ 
&\qquad -
\frac{q^2}{r d} \textbf{F} \left( \tfrac{1}{2} (\delta\theta + \theta_0 - \Theta_0), k \right) -
\frac{2(a - r)}{d} \textbf{F} \left( \tfrac{1}{2} (-\delta\theta - \theta_0 + \Theta_0), k \right) \notag \\
&\qquad -
\frac{d}{r} \textbf{F} \left( \tfrac{1}{2} (\delta\theta - \theta_0 + \Theta_0), k \right) \notag \\
&\qquad +
\frac{2 a q^2}{p^2 C_1} \sin(\delta\theta + \theta_0 - \Theta_0) + 
\frac{2 a q^2}{p^2 C_2} \sin(\delta\theta - \theta_0 + \Theta_0)
\Bigg) \\
a_\theta &= -\frac{G M}{2 \delta\theta r} \left( -\frac{1}{C_1} + \frac{1}{C_2} \right)
\end{align}
\end{subequations}

\begin{figure}[!ht]
\centering
    \includegraphics[width=\linewidth, keepaspectratio]{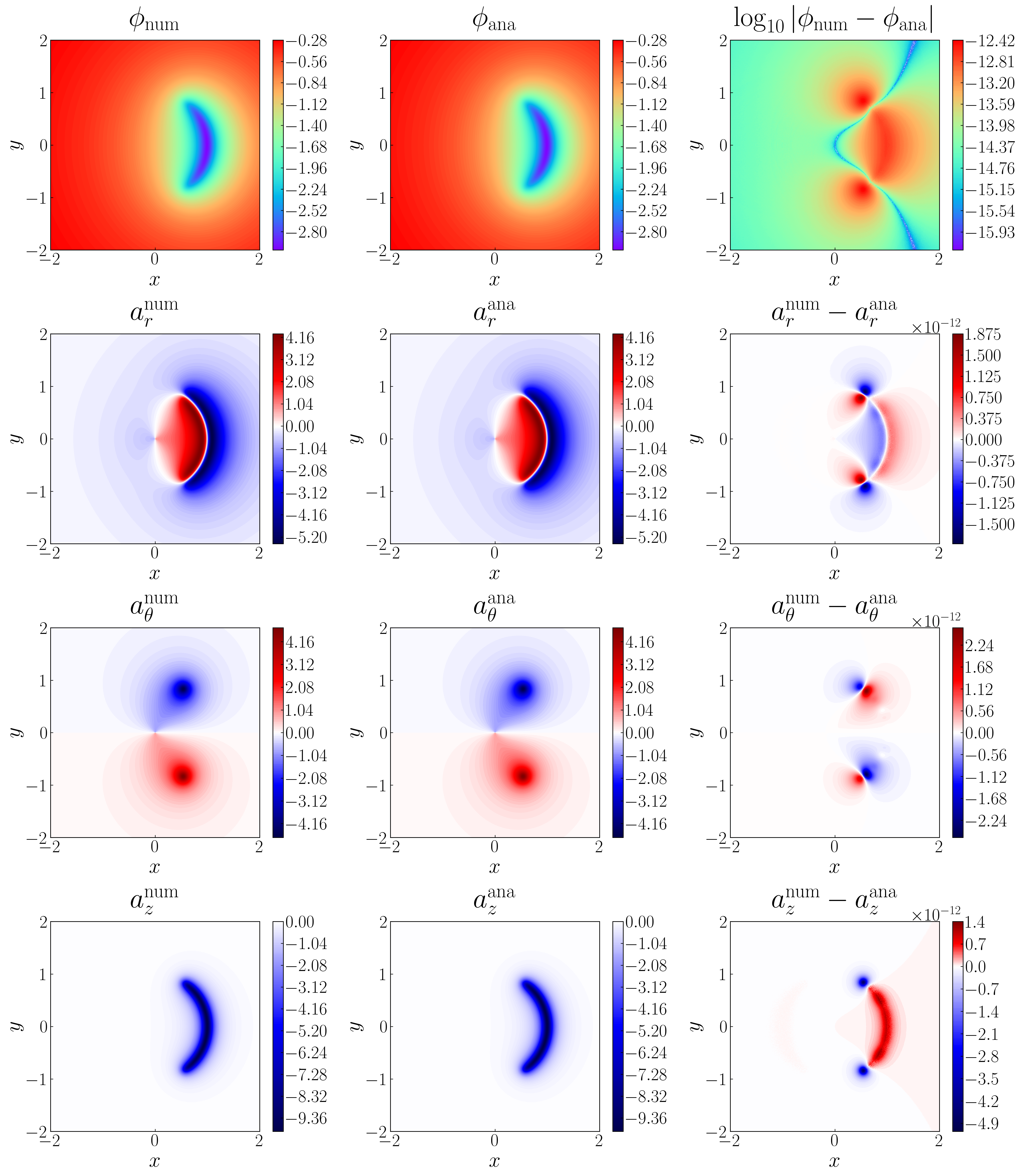}
    \caption{The top left panel shows the result of a numerical integration $\phi_{num}$, the top center panel shows the analytic form $\phi_{ana}$ given by Eq \ref{eq:arcpot} and the top right panel shows the log of their difference.  The agreement is good to machine precision far from the ring and good to $10^{-12}$ near it, due to growing integration error caused by the steepening slope of the potential.  Finally, the acceleration components given by Eq \ref{eq:arc_acc} are shown in the bottom row.  All quantities were computed assuming $G = M = a = 1$,$\delta\theta = 1.0$, $\Theta_0 = 0.0$ and $z_0=0.1$ }
    \label{fig:arccomp}
\end{figure}

Plots of these potentials and accelerations are shown in Fig \ref{fig:arccomp}.  They are compared to a numerical $\phi_{num}$, which is the result of a direct quadrature integration of Eq \ref{eq:arcbase}.  The numerical accelerations are computed with \texttt{JAX} automatic differentiation \citep{Jax_2018}.

\subsection{Limiting Case: Uniform Circular Ring}

In the limiting case of $\delta \theta = \pi$, The expression for a circular arc transforms into that of a ring.  This limiting case has long been described, by \citep{Lass_1983,Fukushima_2010Ring} and others.  In this limit the integral becomes:

\begin{equation}
    V = -\frac{G M}{2\pi } \int_{0}^{2 \pi} \frac{1}{\sqrt{a^2 + r_0^2 + z_0^2 - 2 a r_0 \cos(\theta - \theta_0)}} \, d\theta.
\label{eq:ringbase}
\end{equation}

This integral has a closed form in terms of elliptic integrals

\begin{equation}
    \phi = -\frac{2 G M}{\pi d}  \textbf{K}(k)
\label{eq:ringpot}
\end{equation}

Where $\textbf{K}$ is the complete elliptic integral of the first kind. The corresponding accelerations are given by

\begin{subequations}\label{eq:ring_acc}
\begin{align}
a_z &= -\frac{G M }{\pi p^2} \fracbrac{z_0}{d} \textbf{E}(k) \\
a_r &= \frac{G M}{\pi r_0 d} \left( \fracbrac{q^2}{p^2} \textbf{E}(k) - \textbf{K}(k) \right)\\
a_\theta &= 0.
\end{align}
\end{subequations}

\begin{figure}[!ht]
\centering
    \includegraphics[width=\linewidth, keepaspectratio]{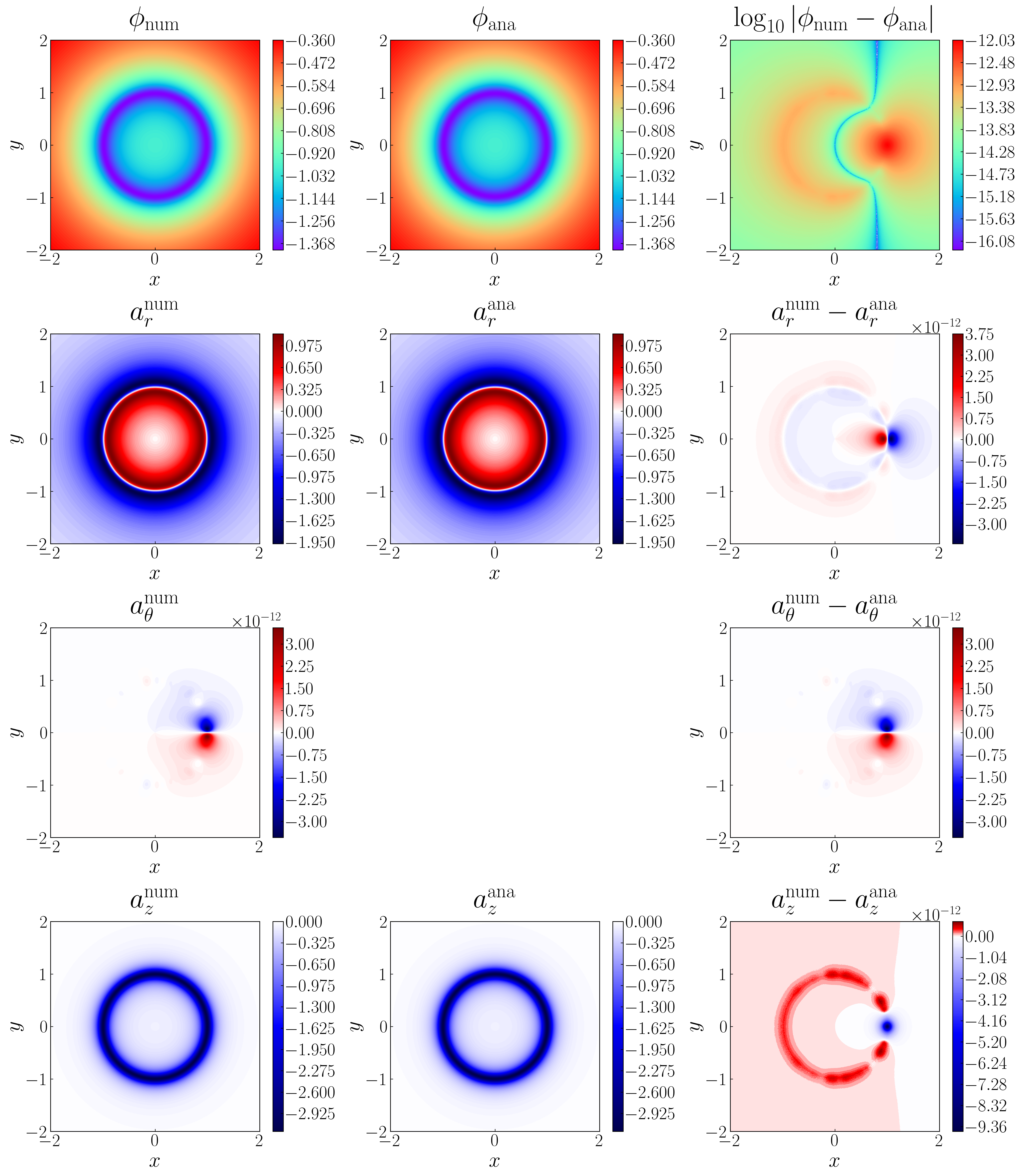}
    \caption{The top left panel shows the result of a numerical integration $\phi_{num}$, the top center panel shows the analytic form $\phi_{ana}$ given by Eq \ref{eq:ringpot} and the top right panel shows the log of their difference.  The agreement is good to machine precision far from the ring and good to $10^{-13}$ near it, due to growing integration error caused by the steepening slope of the potential.  Finally, the acceleration components given by Eq \ref{eq:ring_acc} are shown in the lower rows.  All quantities were computed assuming $G = M = a = 1$,$\delta\theta = 1.0$, $\Theta_0 = 0.0$ and $z_0=0.1$ }
    \label{fig:ringcomp}
\end{figure}

An example of this limiting case are the potentials and accelerations are shown in Fig \ref{fig:ringcomp}.  They are compared to a numerical $\phi_{num}$, which is the result of a direct integration of Eq \ref{eq:ringbase}.

\section{Eccentric Orbit Averaged Arcs}

The simplest case of an eccentric ring is one of uniform density.  The integral is essentially unchanged.

\begin{align}
\phi &= - G \lambda \int_S \frac{r\, d\theta}{|\vec{x}-\vec{r}|} \notag \\
&= \frac{-G M }{\textbf{E}(E_2,e^2) + \textbf{E}(E_1,e^2)} 
\fracbrac{r_0}{a} 
\int_{\Theta_0 - \delta \theta}^{\Theta_0 + \delta \theta} 
\frac{1}{\sqrt{R(\theta)^2 + r_0^2 + z_0^2 - 2 R(\theta) r_0 \cos(\theta - \theta_0)}} \, d\theta
\end{align}

Except here, instead of $R=a$ as in the circular case, $R$ is now a function of $\theta$ such that $R(\theta) = \frac{a (1 - e^2)}{1 + e \cos(\theta - \omega)}$. The linear mass density $\lambda$ is now proportional to the perimeter of a portion of an ellipse, a problem which, famously, can only be expressed in terms of elliptic integrals \citet{AbramowitzStegun1964}.  Since rotational symmetry is lost for any $e \geq 0$, the linear mass density becomes dependent on where the boundary lies with respect to $\omega$. In terms of elliptic integrals, the density is $\lambda = M^{-1} a \left( \textbf{E}(E_2,e^2) + \textbf{E}(E_1,e^2) \right)$ with $E_i = 2 \arctan\left( \sqrt{\frac{1-e}{1+e}}  \tan \fracbrac{\theta_i - \omega}{2}\right)$. 

The essential parts of the integral can be more easily seen by transforming it to the following form

\begin{align}\label{eq:cosform}
V &= \int_{\Phi_{\min}}^{\Phi_{\max}} f(\Phi) \, d\Phi \notag \\
&= \int_{\Phi_{\min}}^{\Phi_{\max}} 
\frac{1 + e \cos \Phi}
{\sqrt{C_1 + \cos(d\Phi - \Phi)\left(C_2 + C_4 \cos \Phi\right) + \cos \Phi \left(C_3 + C_5 \cos \Phi\right)}} \, d\Phi
\end{align}

where $\Phi = \theta - \omega, d\Phi = \theta_0 - \omega$, and the $C_i$ are appropriately defined intermediate variables.  It is now clear that the integral reduces to a linear combination of $1/d$ where $d^2$ is just a polynomial in $\cos{\Phi}$.  This integral can be reduced further via a Weierstrass substitution $t = \tan\frac{\phi}{2}$, which is used to transform $f(\Phi) d\Phi \rightarrow f(t) dt$ . The new integrand becomes 

\begin{equation}\label{eq:uniformecc}
    f(t) dt = \left(2 + e \left(-2 + \frac{4}{1 + t^2}\right) \right) \frac{1}{\sqrt{k_0 + k_1 t + k_2 t^2+ k_3 t^3 + k_4 t^4}}
\end{equation}

with $k_i$ defined as
\begin{subequations}\label{eq:kparams}
\begin{align}
k_0 &= (1 + e)^2 \left( 
a^2 (1 - e)^2 + r_0^2 + z_0^2 + 2a(1 - e) r_0 \cos(d\Phi) 
\right) \\
k_1 &= 4a(1 - e)(1 + e)^2 r_0 \sin(d\Phi) \\
k_2 &= 2(1 - e^2) \left( 
a^2 (1 - e^2) - r_0^2 - z_0^2 - 2ae r_0 \cos(d\Phi) 
\right) \\
k_3 &= -4a(1 - e)^2(1 + e) r_0 \sin(d\Phi) \\
k_4 &= (1 - e)^2 \left(
a^2 (1 + e)^2 + r_0^2 + z_0^2 + 2a(1 + e) r_0 \cos(d\Phi) 
\right)
\end{align}    
\end{subequations}

This transformation comes the cost of having to split the domain of the integration about $\pi$. Therefore, for arbitrary $\Phi_{min}$ and $\Phi_{max}$

\begin{equation}\label{eq:indef}
\int_{\Phi_{\min}}^{\Phi_{\max}} f(\Phi) \, d\Phi =
\begin{cases}
\displaystyle
\int_{\tan\left( \frac{\Phi_{\min}}{2} \right)}^{\tan\left( \frac{\Phi_{\max}}{2} \right)} f(t) \, dt,
& \text{if } \Phi_{\max} \leq \pi \\[12pt]

\displaystyle
\int_{\tan\left( \frac{\Phi_{\min}}{2} \right)}^{\infty} f(t) \, dt
+ \int_{-\infty}^{\tan\left( \frac{\Phi_{\max}}{2} \right)} f(t) \, dt,
& \text{if } \Phi_{\max} > \pi.
\end{cases}
\end{equation}

However, the advantage is to reduce the integral to a known problem - that of the integral of a rational function times the square root of a quartic polynomial.  This is a well known form which can be written in terms of elliptic integrals, by factoring the quartic polynomial and expressing the result in terms of its roots via a Möbius transformation. The result, while dense, can be derived with any computer algebra system. Using \texttt{Mathematica}, we find the indefinite integral can be written as

\begin{equation}
   F(t) = \int f(t) dt = C S(t) E(t).
\end{equation}

With 
\begin{equation}
    C  = -\frac{4}{|r1 - r3| \sqrt{k5}}
\end{equation}

\begin{equation}
S(t) = \frac{\lvert r_{1} - r_{3} \rvert \,(r_{1} - r_{4})(-r_{2} + t)^{2}}
     {(-r_{1} + r_{2})(r_{2} - r_{4})}
\;\sqrt{\frac{(r_{1} - r_{2})^{2}(r_{2} - r_{4})}
              {(r_{1} - r_{3})(r_{1} - r_{4})^{2}(-r_{2} + t)^{4}}}
\end{equation}

\begin{equation}\label{eq:etref}
    E(t) = k_F \boldsymbol{F}(\phi(t),m) +  k_{\Pi1} \boldsymbol{\Pi}(n_1,\phi(t),m) + k_{\Pi2} \boldsymbol{\Pi}(n_2,\phi(t),m)
\end{equation}

Where we've defined the auxiliary quantities : 

\begin{subequations}
\begin{align}
    k_{F}   &= 1 + e\!\left(\frac{2}{1 + r_{2}^{2}} - 1\right) 
    &\qquad 
    s(t)   &= \frac{(r_{2} - r_{4})(-(r_{1} - t))}{(r_{1} - r_{4})(-(r_{2} - t))} \\[6pt]
    k_{\Pi 1} &= \frac{(i e)(r_{1} - r_{2})}{(-i + r_{1})(-i + r_{2})} 
    &\qquad 
    \phi(t) &= \arcsin\!\left(\sqrt{s(t)}\right) \\[6pt]
    k_{\Pi 2} &= -\,\frac{(i e)(r_{1} - r_{2})}{(i + r_{1})(i + r_{2})} 
    &\qquad 
    m      &= -\frac{(r_{2} - r_{3})(r_{1} - r_{4})}{(-r_{1} + r_{3})(r_{2} - r_{4})} \\[6pt]
    n_{1}  &= \frac{(-i + r_{2})(r_{1} - r_{4})}{(-i + r_{1})(r_{2} - r_{4})} 
    &\qquad 
    n_{2}  &= \frac{(i + r_{2})(r_{1} - r_{4})}{(i + r_{1})(r_{2} - r_{4})}.
\end{align}    
\end{subequations}

Here $\boldsymbol{\Pi}$ is the Legendre incomplete elliptic integrals of the third kind, and $r_i$ are the roots of the quartic polynomial in the denominator in Eq \ref{eq:uniformecc}. These roots can be computed brute-force through the full quartic formula, but can also be done without complex arithmetic via the well-known Ferrari and Cardano decompositions \citep{AbramowitzStegun1964,Cardano_1968}.  Since the polynomial is always greater than zero, these roots are complex and come in conjugate pairs, we choose indices such that  $r_1 = A_1 + i B_1 = r_2^*$ and $r_3 = A_2 + i B_2 = r_4^*$.  While the conjugacy conditions guarantee that certain variables are always real (e.g $C$ and $m$), the elliptic functions of the --- generally complex --- arguments yield complex results with nonzero real and imaginary parts, and so are fully complex functions of $t$.  The $t$ dependence of $F(t)$ is carried through $S(t)$ and $s(t)$, however the behavior of $S(t)$ is very simple as it is either $-1$ or $1$ depending on t, so the dependence is equivalent to a sign change. The rest of the variables are also generally complex, with $k_{\pi 1}$ and $k_{\pi 2}$ being symmetric about the imaginary axis, with the same imaginary part but opposite signed real parts.

\begin{figure}[!ht]
\centering
    \includegraphics[width=\linewidth, keepaspectratio]{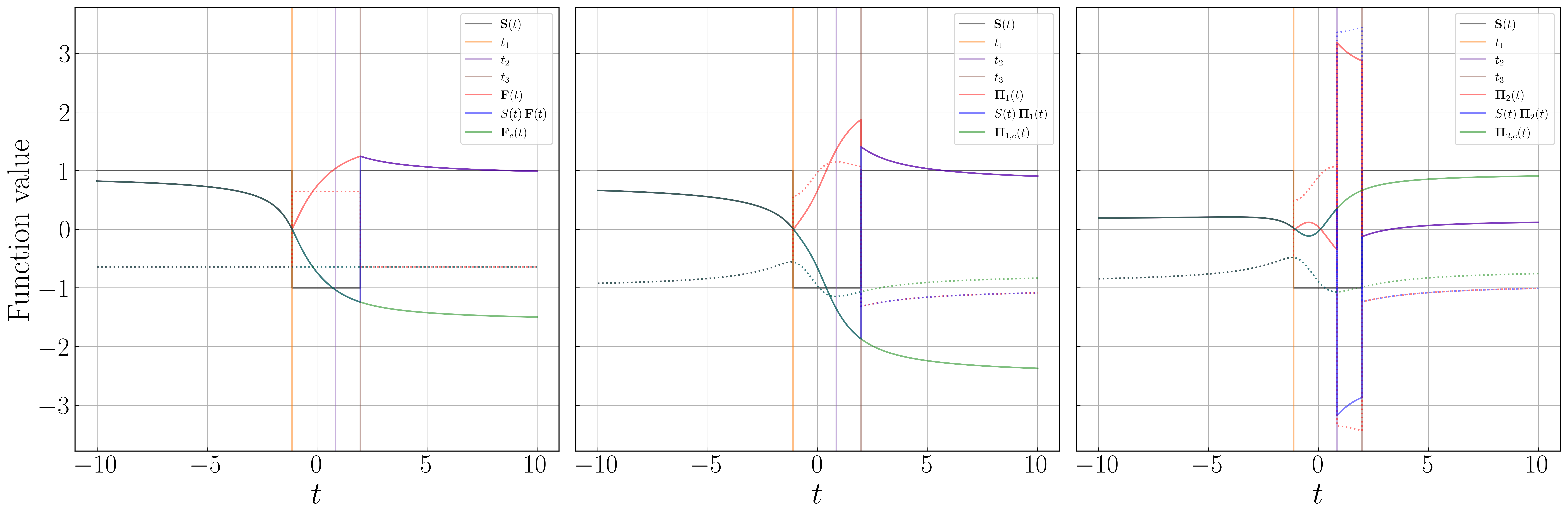}
    \caption{We show an example of the discontinuous behavior of the $\boldsymbol{F}$, $\boldsymbol{\Pi_1}$ and $\boldsymbol{\Pi_2}$ functions described in Eq \ref{eq:etref}.   In each case, we show the function itself in red, the function times $S(t)$ in blue, and the continued, patched function in green.  All quantities were computed assuming $G = M = a = 1$, $r_0=0.5$, $e=0.5$, $\omega = 2.0$, $\Theta_0 = 0.0$ and $z_0=0.1$ }
    \label{fig:continunity}
\end{figure}

Sadly, the branch cuts and other discontinuities in these functions displace the real part of the function and cause the imaginary part to be piecewise constant (rather than fully constant, which would cancel in Eq \ref{eq:indef} when the bounds are evaluated) . Hence, while the indefinite integral is known, care must be taken when evaluating the integrand to compute the definite integral, as particular choices of the integration bounds lead to an imaginary residual and an incorrect real result, explicitly $\int_a^b f(t) \neq F(b) - F(a)$. This issue can be seen graphically in Fig \ref{fig:continunity} for the $\boldsymbol{F}$ and $\boldsymbol{\Pi}$ functions.  While these properties could be easily overcome numerically, by checking if a discontinuities is crossed and shifting the result accordingly, their presence reduces the analytic utility of the result and increases the computational cost of evaluation. Consequently, we take the approach of manually finding and analytically patch the offsets. To this end, we wish to derive a continuous version of the antiderivative, which we call $F_c(t)$.

Since the sum of continuous functions is continuous, it is most convenient to derive $F_c(t)$ by defining continuous versions of the individual elliptic integrals (which we will denote with $\boldsymbol{F_c}$ and $\boldsymbol{\Pi_c}$). We can then transform $F(t) \rightarrow F_c(t)$ by letting $S(t)\boldsymbol{F} \rightarrow \boldsymbol{F_c}$ and $S(t)\boldsymbol{\Pi} \rightarrow \boldsymbol{\Pi_c}$.  In this way we can absorb the piecewise behavior of $S(t)$ and the other discontinuities into a single function. To do this we must understand the locations of the discontinuities and their magnitudes, in addition to the behavior of the $S(t)$ function.  We begin with $S(t)$. 

Since the branch cut for the square root lies on the negative imaginary axis, we seek values of $t$ such that  

\begin{equation}
\begin{aligned}
\Im\!\left[
\frac{(r_{1} - r_{2})^{2}(r_{2} - r_{4})}{(r_{1} - r_{3})(r_{1} - r_{4})^{2}(-r_{2} + t)^{4}}
\right] &= 0,\\[6pt]
\Re\!\left[
\frac{(r_{1} - r_{2})^{2}(r_{2} - r_{4})}{(r_{1} - r_{3})(r_{1} - r_{4})^{2}(-r_{2} + t)^{4}}
\right] &< 0.
\end{aligned}
\end{equation}

The two solutions can be expressed in terms of $A_i$ and $B_i$ as

\begin{subequations}
\begin{align}
D &:= 
\sqrt{
\bigl((A_{1} - A_{2})^{2} + B_{1}^{2}\bigr)^{2}
+ 2 (A_{1} - A_{2} - B_{1})(A_{1} - A_{2} + B_{1}) B_{2}^{2}
+ B_{2}^{4}
} \\[6pt]
t_{1} &:= -\frac{-A_{1}^{2} + A_{2}^{2} - B_{1}^{2} + B_{2}^{2} - D}
{2 (A_{1} - A_{2})} \\[6pt]
t_{3} &:= -\frac{-A_{1}^{2} + A_{2}^{2} - B_{1}^{2} + B_{2}^{2} + D}
{2 (A_{1} - A_{2})}.
\end{align}
\end{subequations}

Another branch cut can occur in $\phi(t)$ due to $s(t)$ crossing a branch cut. If we solve for the analogous condition we find it is not independent, but occurs at the same location as our $t_{3}$ from earlier and is therefore already accounted for.  Finally, there can be one last discontinuity where the $\boldsymbol{\Pi}$ function crosses a pole. This occurs where $n s(t) = 0$.  Solving for this t we get a unique result 

\begin{equation}
    t_2 := \frac{-1 + A_1^2 + B_1^2 - \sqrt{(1 + A_1^2)^2 + 2(-1 + A_1^2)B_1^2 + B_1^4}}{2A_1}
\end{equation}

which we define to be $t_2$.  We've chosen our indexes such that $t_1 < t_2 < t_3$.  Therefore, the function then splits into 4 piecewise parts over the domains $t<t_1$, $t_1<t<t_2$, $t_2<t<t_3$ and $t_3 < t$.  

Since the definite integral should be continuous, we now must derive the proper corrections for each part. Since we use a diverse set of functions in this work, we provide corrections corrections for the $\boldsymbol{\Pi}$, $\boldsymbol{F}$ and $\boldsymbol{E}$ elliptic functions and list all of them here. 

\begin{subequations}
\begin{align}
\Delta_{\Pi,B,i}(n_i,s(t)^2,m)
&=
\\[4pt]
&\quad -\,\frac{2n_i\,s(t)^{3}}{3}\;
   \boldsymbol{R_J}\!\bigl((m\!-\!1)s(t)^2,0,ms(t)^2,(m\!-\!n_i)s(t)^2\bigr)
\\[4pt]
&\quad-\;
   2\,s(t)\;
   \Re\!\bigl[
     \boldsymbol{R_F}\!\bigl(1-s(t)^2,1-m s(t)^2,1\bigr)
   \bigr],
\\[10pt]
\Delta_{\Pi,P,i}(n_i,m)
&= i\,\pi\,\sqrt{\frac{n_i}{(n_i-1)\bigl(n_i-m\bigr)}},
\\[10pt]
\Delta_F
&= -\,2\,\Re\!\bigl[\,\boldsymbol{K}(m)\,\bigr],
\\[4pt]
\Delta_E
&= -\,2\,\Re\!\bigl[\,\boldsymbol{E}(m)\,\bigr].
\end{align}
\end{subequations}

We note the $\Delta_{\Pi,B,i}$ is very close to a complete $\boldsymbol{\Pi}$ function, however the real part of the $\boldsymbol{R_F}$ function ensures no a simple form in terms of $\boldsymbol{\Pi}$ is possible. For derivations of the $\boldsymbol{\Pi}$ jumps we direct readers to the appendix.

\begin{table}[h!]
\centering
\small
\renewcommand{\arraystretch}{1.3}
\begin{tabular}{c|c|c|c}
\hline
\textbf{$t$ interval} 
& $\mathbf{F}_c(t)$ 
& $\boldsymbol{\Pi}_{1,c}(t)$ 
& $\boldsymbol{\Pi}_{2,c}(t)$ \\
\hline
$ t < t_1 $ 
& $ \mathbf{F} $ 
& $ \boldsymbol{\Pi}_1 $ 
& $ \boldsymbol{\Pi}_2 $ \\

$ t_1 \le t < t_2 $ 
& $ -\mathbf{F} $ 
& $ -\boldsymbol{\Pi}_1 $ 
& $ -\boldsymbol{\Pi}_2 $ \\

$ t_2 \le t < t_3 $ 
& $ -\mathbf{F} $ 
& $ -\boldsymbol{\Pi}_1 $ 
& $ -\boldsymbol{\Pi}_2 - \Delta_{\Pi,B,2} $ \\

$ t \ge t_3 $ 
& $ \mathbf{F} + \Delta_F $ 
& $ \boldsymbol{\Pi}_1 + \Delta_{\Pi,B,1} $ 
& $ \boldsymbol{\Pi}_2 + \Delta_{\Pi,P,i} - \Delta_{\Pi,B,2}$ \\
\hline
\end{tabular}
\caption{
Piecewise definitions of $\mathbf{F}_c(t)$, $\boldsymbol{\Pi}_{1,c}(t)$, and $\boldsymbol{\Pi}_{2,c}(t)$.
Here, $\mathbf{F} \equiv \mathbf{F}\!\big(\phi(t)\mid m_F\big)$,
$\boldsymbol{\Pi}_1 \equiv \boldsymbol{\Pi}\!\big(n;\,\phi(t)\mid m_{\Pi,1}\big)$,
and $\boldsymbol{\Pi}_2 \equiv \boldsymbol{\Pi}\!\big(n;\,\phi(t)\mid m_{\Pi,2}\big)$.
}
\label{tab:table1}
\end{table}

Finally, we can explicitly represent $F_c(t)$ as

\begin{equation}\label{eq:Fc}
    F_c(t) = C(k_F \boldsymbol{F_c}(\phi(t),m) +  k_{\Pi1} \boldsymbol{\Pi_{1,c}}(n_1,\phi(t),m) + k_{\Pi2} \boldsymbol{\Pi_{2,c}}(n_2,\phi(t),m))
\end{equation}
where we've defined $\boldsymbol{F_c}$,$\boldsymbol{\Pi_{1,c}}$,$\boldsymbol{\Pi_{2,c}}$ in Table \ref{tab:table1}.

Due to the Weierstrass substitution, if we wish to evaluate the function we must do so at the improper bounds $t \rightarrow \pm \infty$.  Since we do not wish to proceed numerically, it would be useful to have analytic expressions for these limits. Since after the elimination of $S(t)$ our function was only $t$ dependent through  $\phi(t)$, we need only to take its limits, both of which are identical

\begin{equation}
  \lim_{t \to \pm\infty} \phi(t)
  \;=\;
  \arcsin\!\left(
    \sqrt{\tfrac{\,r_2 - r_4\,}{\,r_1 - r_4\,}}
  \right)
  \;=:\;\phi^*.
\end{equation}

\begin{figure}[!ht]
\centering
    \includegraphics[width=\linewidth, keepaspectratio]{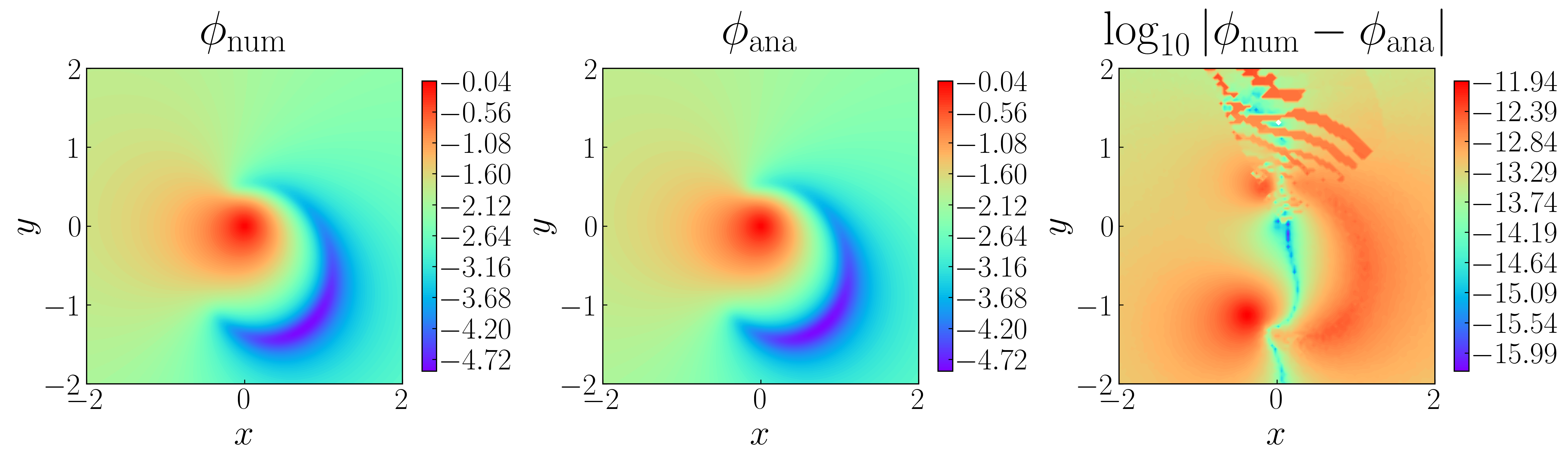}
    \caption{The left panel shows the result of a numerical integration $\phi_{num}$, the center panel shows the analytic form $\phi_{ana}$ given by Eqs \ref{eq:lim1_minus},\ref{eq:lim1_plus},\ref{eq:Fc},\ref{eq:indef}  and the right panel shows the log of their difference.   All quantities were computed assuming $G = M = a = 1$,$\delta\theta = 1.9$, $\Theta_0 = 0.0$, $e=0.5$, $\omega=2.0$ and $z_0=0.1$  }
    \label{fig:eccarc}
\end{figure}

Hence,

\begin{equation}\label{eq:lim1_minus}
  \lim_{t \to -\infty} F_c(t) =  C\Big[
      k_F\,\mathbf{F}\!\big(\phi^*\mid m\big)
    + k_{\Pi1}\,\boldsymbol{\Pi}\!\big(n_1;\,\phi^*\mid m\big)
    + k_{\Pi2}\,\boldsymbol{\Pi}\!\big(n_2;\,\phi^*\mid m\big)
  \Big],
\end{equation}

\begin{align}\label{eq:lim1_plus}
  \lim_{t \to +\infty} F_c(t) &= C\Big[
      k_F\,\mathbf{F}\!\big(\phi^*\mid m\big)
    + k_{\Pi1}\,\boldsymbol{\Pi}\!\big(n_1;\,\phi^*\mid m\big)
    + k_{\Pi2}\,\boldsymbol{\Pi}\!\big(n_2;\,\phi^*\mid m\big) \notag \\[3pt]
  &\qquad\;\;+\;
      k_F\,\Delta_F
    + k_{\Pi1}\,\Delta_{\Pi,B,1}
    + k_{\Pi2}\,\Delta_{\Pi,P,i}
    - k_{\Pi2}\,\Delta_{\Pi,B,2}
  \Big].
\end{align}

we now have everything we need to derive the potential for any point in space, with any given orbital parameters. A test of these equations is preformed in Fig \ref{fig:eccarc}, where we compare a numerically integrated potential to the analytic prediction using Eqs \ref{eq:lim1_minus},\ref{eq:lim1_plus},\ref{eq:Fc},\ref{eq:indef}.

\subsection{Limiting Case, Eccentric Orbit Averaged Ring}

While we have successfully derived an analytic potential for the general orbit averaged case, the most interesting case is that which occurs when $\Phi_{min} =0$ and $\Phi_{max}= 2\pi$.  In this limit our integral integral in Eq \ref{eq:indef}, always takes the second branch.  Since $\Phi_{max} = \Phi_{min}$ the two terms collapse and we are left with only the difference between the two improper limits $\lim_{t \to +\infty} F_c(t) -  \lim_{t \to -\infty} F_c(t)$.  Here again, most terms cancel, leaving the entire integral expressible solely in terms of the corrections we derived for the individual pieces.  

\begin{equation}
V = C(k_F \,\Delta_F + k_{\Pi1}\,\Delta_{\Pi,B,1} + k_{\Pi2}\,\Delta_{\Pi,P,i} - k_{\Pi2}\,\Delta_{\Pi,B,2})    
\end{equation}

\begin{figure}[!ht]
\centering
    \includegraphics[width=\linewidth, keepaspectratio]{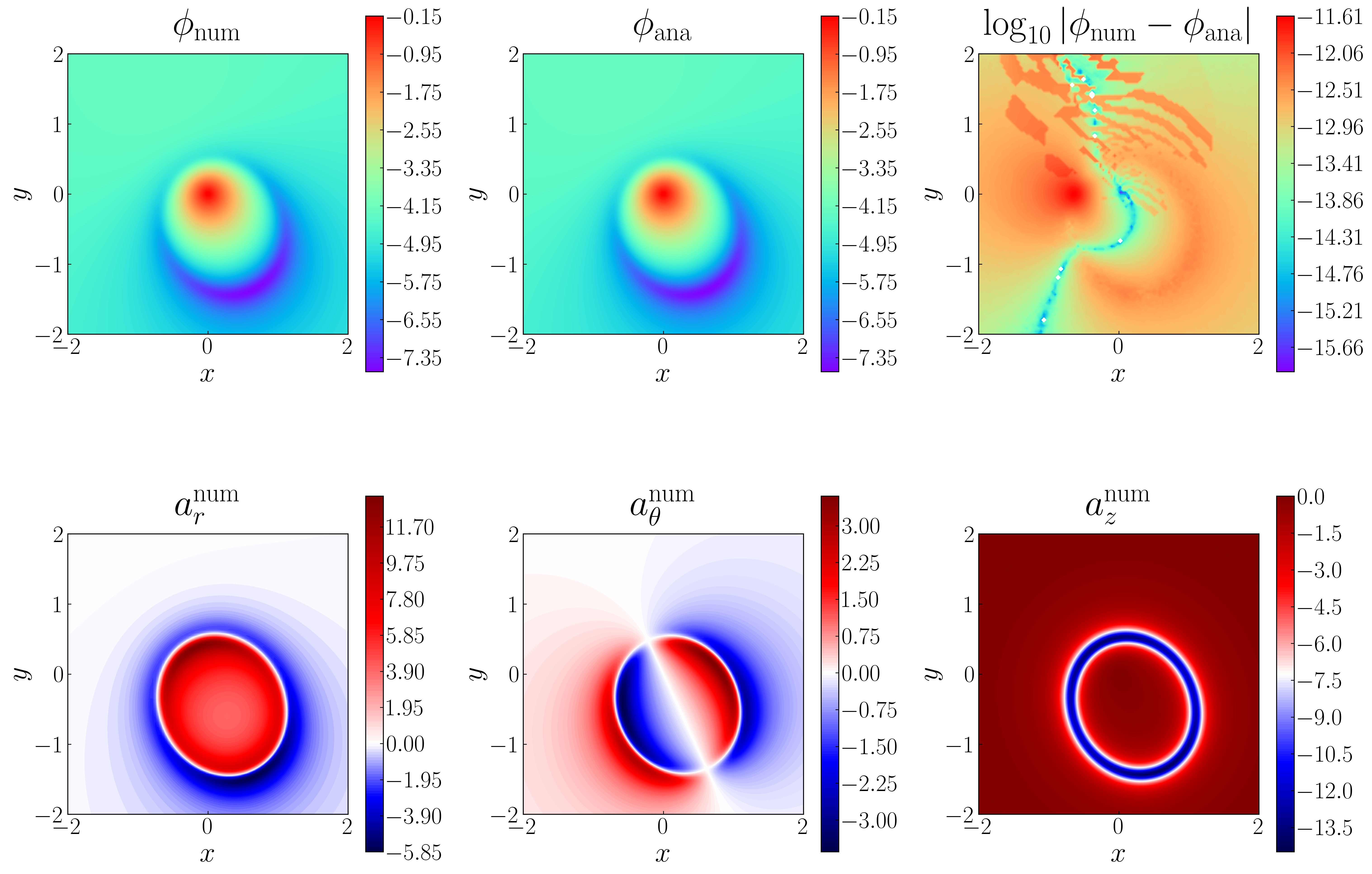}
    \caption{The top left panel shows the result of a numerical integration $\phi_{num}$, the top center panel shows the analytic form $\phi_{ana}$  and the top right panel shows the log of their difference. The lower rows shows the accelerations in $r$, $\theta$,$z$ directions.   All quantities were computed assuming $G = M = a = 1$,$\delta\theta = \pi$, $\Theta_0 = 0.0$, $e=0.5$, $\omega=2.0$ and $z_0=0.1$ }
    \label{fig:eccring}
\end{figure}

A test of the potential and the corresponding accelerations are shown  in Fig \ref{fig:eccring}.  While the potential looks qualitatively like what one might expect from an eccentric orbit, a closer examination shows several differences that can be seen in the accelerations.  First the $a_r$ acceleration is strongest outward (most positive) just inside the periapsis whereas it is strongest inward (most negative) just outside the apoapsis. Hence the inward and outward accelerations are anti-aligned, which is the opposite of the time-averaged case.  In addition, the accelerations in $a_z$ is completely uniform, while the sign of $a_{\theta}$ depends only on your location with respect to the $\omega$. Both of these features are absent from a time averaged ring.  While the uniform density ellipse my serve as an appropriate model small eccentricities, becomes an increasingly poor approximation for large eccentricities.  This form is useful for demonstrating the methodology and for qualitative modeling, but is of limited dynamical use.

\section{Eccentric Time Averaged Rings}

Having derived the expression for a simple orbit averaged arc and ring, we now turn our attention to the most dynamically relevant case, that of a time averaged mass distribution appropriate for a point massive body in a Keplerian orbit.  The form of the integral is similar to that of the eccentric uniform ring, except the mass distribution must now accommodate for the fact that an orbiting body spends most of it's time near apoapsis.  The integral takes the form

\begin{align}
\phi 
&= - G M \int_S \frac{\lambda(\theta)\, r\, d\theta}{|\vec{x}-\vec{r}|} \notag \\[4pt]
&= \frac{-G M (1-e^2)^{3/2}}{2 \pi} 
   \int_{\Theta_0 - \delta \theta}^{\Theta_0 + \delta \theta} 
   \frac{(1 + e \cos(\theta - \omega))^2}
   {\sqrt{R(\theta)^2 + r^2 + z^2 - 2 R(\theta) r \cos(\theta - \theta_0)}} 
   \, d\theta
\end{align}

Here we retain the definition that $R(\theta) = \frac{a (1 - e^2)}{1 + e \cos(\theta - \omega)}$, as in the eccentric case but choose the linear mass density to be $\lambda(\theta) = \frac{M (1-e^2)^{3/2} }{2 \pi (1+e \cos(\theta-\omega)^2)}$.  The essential form of the integral is essentially identical to the uniform case and uses all the same constants, the only difference is the additional factor of $1+ e \cos(\Phi)$ appearing in the denominator instead of the numerator in the equivalent of Eq \ref{eq:cosform}. Applying the Wierstrass transformation yields the following (subject to the same caveat as in Eq \ref{eq:indef})

\begin{equation}\label{eq:indefkepler}
    f(t) dt = \left( \frac{2(1 + t^{2})}{1 + e + t^{2} - e t^{2}} \right) \frac{1}{\sqrt{k_0 + k_1 t + k_2 t^2+ k_3 t^3 + k_4 t^4}}.
\end{equation}

The solution in terms of elliptic integrals can be written in the same form

\begin{equation}
    F(t) = c (k_F \boldsymbol{F}(\phi(t),m) +  k_{\Pi1} \boldsymbol{\Pi}(n_1,\phi(t),m) + k_{\Pi2} \boldsymbol{\Pi}(n_2,\phi(t),m)).
\end{equation}

However, it uses somewhat different algebraic substitutions. Where we defined $\tau_1 = \sqrt{1 + e}$ and $\tau_2 = \sqrt{e-1}$

\begin{subequations}
\begin{align}
k_{F} &= \tau_{2}^{2}\tau_{1}\bigl(-\tau_{1}^{2}\!+\!\tau_{2}^{2}r_{1}^{2}\bigr)(1\!+\!r_{2}^{2}) \\[6pt]
k_{\Pi 1} &= -e(r_{1}\!-\!r_{2})\Bigl(
    \tau_{2}\!-\!\tau_{1}r_{1}\!-\!\tau_{1}r_{2}\!+\!\tau_{2}r_{1}r_{2}
    + e(\tau_{2}\!+\!\tau_{1}r_{2}\!+\!r_{1}(\tau_{1}\!+\!\tau_{2}r_{2}))
\Bigr) \\[6pt]
k_{\Pi 2} &= e(r_{1}\!-\!r_{2})\Bigl(
    \tau_{2}\!+\!\tau_{1}r_{1}\!+\!\tau_{1}r_{2}\!-\!\tau_{2}r_{1}r_{2}
    + e(\tau_{2}\!-\!\tau_{1}r_{1}\!-\!\tau_{1}r_{2}\!+\!\tau_{2}r_{1}r_{2})
\Bigr) \\[6pt]
s(t) &= \frac{(r_{2}\!-\!r_{4})(-(r_{1}\!-\!t))}{(r_{1}\!-\!r_{4})(-(r_{2}\!-\!t))} \\[6pt]
\phi(t) &= \arcsin\!\left(\sqrt{s(t)}\right) \\[6pt]
m &= -\frac{(r_{2}\!-\!r_{3})(r_{1}\!-\!r_{4})}{(-r_{1}\!+\!r_{3})(r_{2}\!-\!r_{4})} \\[6pt]
n_{1} &= \frac{(\tau_{1}\!-\!\tau_{2}r_{2})(r_{1}\!-\!r_{4})}{(\tau_{1}\!-\!\tau_{2}r_{1})(r_{2}\!-\!r_{4})} \\[6pt]
n_{2} &= \frac{(\tau_{1}\!+\!\tau_{2}r_{2})(r_{1}\!-\!r_{4})}{(\tau_{1}\!+\!\tau_{2}r_{1})(r_{2}\!-\!r_{4})}
\end{align}
\end{subequations}

\begin{equation}
C \;=\; -\,\frac{4}{\tau_{2}^{2}\,\tau_{1}\,
\bigl(-\tau_{1}^{2}+\tau_{2}^{2}r_{1}^{2}\bigr)\,
\bigl(-\tau_{1}^{2}+\tau_{2}^{2}r_{2}^{2}\bigr)\,
\sqrt{(r_{1}-r_{3})(r_{2}-r_{4})}\,\sqrt{k_{5}}}\,.    
\end{equation}

\begin{equation}
    S(t) \;=\; 
\frac{\dfrac{r_{1}-r_{4}}{(r_{1}-t)(r_{4}-t)}}%
{\sqrt{\dfrac{(r_{1}-r_{4})^{2}}{(r_{1}-t)^{2}(r_{4}-t)^{2}}}}.
\end{equation}

The corresponding splits are

\begin{align}
D &:= 
\sqrt{
   \bigl((A_{1} - A_{2})^{2} + B_{1}^{2}\bigr)^{2}
   + 2 (A_{1} - A_{2} - B_{1})(A_{1} - A_{2} + B_{1}) B_{2}^{2}
   + B_{2}^{4}
}
\\[6pt]
t_{1} &:=
-\frac{-A_{1}^{2} + A_{2}^{2} - B_{1}^{2} + B_{2}^{2} - D}
{2 (A_{1} - A_{2})}
\\[6pt]
t_{2} &:=
\frac{
   \bigl(A_{1}^{2} + B_{1}^{2}\bigr)\,\lvert \tau_{2}^{2}\rvert - \lvert \tau_{1}^{2}\rvert
   -
   \sqrt{
      \left(\bigl(A_{1}^{2} + B_{1}^{2}\bigr)\,\lvert \tau_{2}^{2}\rvert - \lvert \tau_{1}^{2}\rvert\right)^{2}
      + 4 A_{1}^{2}\,\lvert \tau_{2}^{2}\rvert\,\lvert \tau_{1}^{2}\rvert
   }
}{
   2 A_{1}\,\lvert \tau_{2}^{2}\rvert
}
\\[6pt]
t_{3} &:=
-\frac{-A_{1}^{2} + A_{2}^{2} - B_{1}^{2} + B_{2}^{2} + D}
{2 (A_{1} - A_{2})}.
\end{align}

Studying these splits shows a critical difference with the uniform ellipse.  Unlike in the case of a uniform ring, we are not guaranteed that $t_2 < t_3$, hence we must have two ways of enforcing continuity. One valid for $t_2 < t_3$ and the other for the $t_3>t_2$ case. The appropriate continuations are shown in Table \ref{tab:table2} for the $t_2 < t_3$ case and in Table \ref{tab:table3} for the $t_3 < t_2$ case. 

\begin{table}[h!]
\centering
\small
\renewcommand{\arraystretch}{1.3}
\begin{tabular}{c|c|c|c}
\hline
\textbf{$t$ interval}
& $\mathbf{F}_c(t)$
& $\boldsymbol{\Pi}_{1,c}(t)$
& $\boldsymbol{\Pi}_{2,c}(t)$ \\
\hline
$ t < t_1 $
& $ \boldsymbol{F} $
& $ \boldsymbol{\Pi}_1 + \Delta_{\Pi,P,1} $
& $ \boldsymbol{\Pi}_2 $ \\

$ t_1 \le t < t_2 $
& $ -\boldsymbol{F} $
& $ -\boldsymbol{\Pi}_1 + \Delta_{\Pi,P,1} $
& $ -\boldsymbol{\Pi}_2 $ \\

$ t_2 \le t < t_3 $
& $ \boldsymbol{F} + \Delta_F $
& $ \boldsymbol{\Pi}_1 + \Delta_{\Pi,B,1} + \Delta_{\Pi,P,1} $
& $ \boldsymbol{\Pi}_2 + \Delta_{\Pi,B,2} $ \\

$ t \ge t_3 $
& $ \boldsymbol{F} + \Delta_F $
& $ \boldsymbol{\Pi}_1 + \Delta_{\Pi,B,1} + 2\,\Delta_{\Pi,P,1} $
& $ \boldsymbol{\Pi}_2 + \Delta_{\Pi,B,2} $ \\
\hline
\end{tabular}
\caption{
Continuation for the case $t_2 < t_3$.
Here, $\boldsymbol{F} \equiv \boldsymbol{F}\!\big(\phi(t)\mid m_F\big)$,
$\boldsymbol{\Pi}_1 \equiv \boldsymbol{\Pi}\!\big(n_1;\phi(t)\mid m_{\Pi,1}\big)$,
and $\boldsymbol{\Pi}_2 \equiv \boldsymbol{\Pi}\!\big(n_2;\phi(t)\mid m_{\Pi,2}\big)$.}
\label{tab:table2}
\end{table}

\begin{table}[h!]
\centering
\small
\renewcommand{\arraystretch}{1.3}
\begin{tabular}{c|c|c|c}
\hline
\textbf{$t$ interval}
& $\mathbf{F}_c(t)$
& $\boldsymbol{\Pi}_{1,c}(t)$
& $\boldsymbol{\Pi}_{2,c}(t)$ \\
\hline
$ t < t_1 $
& $ \boldsymbol{F} $
& $ \boldsymbol{\Pi}_1 + \Delta_{\Pi,P,1} $
& $ \boldsymbol{\Pi}_2 $ \\

$ t_1 \le t < t_3 $
& $ -\boldsymbol{F} $
& $ -\boldsymbol{\Pi}_1 + \Delta_{\Pi,P,1} $
& $ -\boldsymbol{\Pi}_2 $ \\

$ t_3 \le t < t_2 $
& $ -\boldsymbol{F} $
& $ -\boldsymbol{\Pi}_1 $
& $ -\boldsymbol{\Pi}_2 $ \\

$ t \ge t_2 $
& $ \boldsymbol{F} + \Delta_F $
& $ \boldsymbol{\Pi}_1 + \Delta_{\Pi,B,1} $
& $ \boldsymbol{\Pi}_2 + \Delta_{\Pi,B,2} $ \\
\hline
\end{tabular}
\caption{
Continuation for the case $t_3 < t_2$.
As before,
$\boldsymbol{F} \equiv \boldsymbol{F}\!\big(\phi(t)\mid m_F\big)$,
$\boldsymbol{\Pi}_1 \equiv \boldsymbol{\Pi}\!\big(n_1;\phi(t)\mid m_{\Pi,1}\big)$,
and $\boldsymbol{\Pi}_2 \equiv \boldsymbol{\Pi}\!\big(n_2;\phi(t)\mid m_{\Pi,2}\big)$.}
\label{tab:table3}
\end{table}

Finally, we must work out the corresponding limits.  Like on our previous case we obtain the same limit of $\phi(t)$ for both bounds

\begin{equation}
  \lim_{t \to \pm\infty} \phi(t)
  \;=\;
  \arcsin\!\left(
    \sqrt{\tfrac{\,r_2 - r_4\,}{\,r_1 - r_4\,}}
  \right)
  \;=:\;\phi^*.
\end{equation}

Hence,

\begin{equation}
\begin{aligned}
\lim_{t \to -\infty} F(t)
&= C\Big[
      k_F\,\boldsymbol{F_c}\!\big(\phi^*\!\mid m\big)
    + k_{\Pi1}\!\Big(
         \boldsymbol{\Pi_{1,c}}\!\big(n_1;\phi^*\!\mid m\big)
         + \Delta_{\Pi,P,1}
       \Big) \\[-0.2em]
&\qquad\quad
    + k_{\Pi2}\,\boldsymbol{\Pi_{2,c}}\!\big(n_2;\phi^*\!\mid m\big)
  \Big],
\end{aligned}
\end{equation}

For the case \(t_2 < t_3\),
\begin{align}
\lim_{t \to +\infty} F(t)
&= C\Big[
      k_F\!\left(\boldsymbol{F_c}(\phi^*\!\mid m) + \Delta_F\right) \notag\\[2pt]
&\qquad
    + k_{\Pi1}\!\left(\boldsymbol{\Pi_{1,c}}(n_1;\phi^*\!\mid m)
                     + \Delta_{\Pi,B,1}
                     + 2\,\Delta_{\Pi,P,1}\right) \notag\\[2pt]
&\qquad
    + k_{\Pi2}\!\left(\boldsymbol{\Pi_{2,c}}(n_2;\phi^*\!\mid m)
                     + \Delta_{\Pi,B,2}\right)
   \Big].
\end{align}

For the case \(t_3 < t_2\),
\begin{align}
\lim_{t \to +\infty} F(t)
&= C\Big[
      k_F\!\left(\boldsymbol{F_c}(\phi^*\!\mid m) + \Delta_F\right) \notag\\[2pt]
&\qquad
    + k_{\Pi1}\!\left(\boldsymbol{\Pi_{1,c}}(n_1;\phi^*\!\mid m)
                     + \Delta_{\Pi,B,1}\right) \notag\\[2pt]
&\qquad
    + k_{\Pi2}\!\left(\boldsymbol{\Pi_{2,c}}(n_2;\phi^*\!\mid m)
                     + \Delta_{\Pi,B,2}\right)
   \Big].
\end{align}

The corresponding potential is:

\begin{equation}
V \;=\; C\Big[\,k_F\,\Delta_F \;+\; k_{\Pi1}\big(\Delta_{\Pi,B,1} + \Delta_{\Pi,P,1}\big) \;+\; k_{\Pi2}\,\Delta_{\Pi,B,2}\,\Big].
\end{equation}

for $t_2<t_3$, and 

\begin{equation}
V \;=\; C\Big[\,k_F\,\Delta_F \;+\; k_{\Pi1}\big(\Delta_{\Pi,B,1} - \Delta_{\Pi,P,1}\big) \;+\; k_{\Pi2}\,\Delta_{\Pi,B,2}\,\Big].
\end{equation}

for $t_3<t_2$. These two expression differ only by the sign of $\Delta_{\Pi,P,1}$.

\section{Eccentric Time Averaged Accelerations}

Now we must derive the accelerations.  There are few cases where elliptic time-averaged arcs would be useful in astrophysics, since it runs against the assumption that the massive body completes a full Keplerian orbit. Given this, we instead provide explicit accelerations only the case where $d\phi = \pi$, where we have a full orbit-averaged eccentric ring.  

To preform this operation, it may be tempting to simply apply the chain rule recursively to the expressions found previously, unfortunately this procedure will produce many complex terms which would be difficult to cancel and simplify.  Instead, we take the approach of taking the gradient of potential indirectly.  Since the potential can be represented in terms of the indefinite integral $F(t)$, it is simpler to take the gradient of this function and integrate the result. The derivatives of Eq \ref{eq:indefkepler} can be written as
\begin{equation}
\frac{\partial F(t)}{\partial p}  =\,\frac{(1+t^2)\,\displaystyle\sum_{i=1}^{5}\frac{\partial k_i}{\partial p}\,t^{\,i-1}}
{(1+e+t^2-e t^2)\,\left(\displaystyle\sum_{i=1}^{5} k_i\,t^{\,i-1}\right)^{3/2}},    
\end{equation}

where $p$ is our parameter ($p\in\{r_0,\theta_0,z_0\}$). The derivatives $k_i$ are are shown in Table \ref{tab:ki-derivatives}.

\begin{table}[h!]
\centering
\small
\renewcommand{\arraystretch}{1.3}
\begin{tabular}{c|c|c|c}
\textbf{$k_i$} 
& $\displaystyle \frac{\partial k_i}{\partial r_0}$ 
& $\displaystyle \frac{\partial k_i}{\partial \theta_0}$ 
& $\displaystyle \frac{\partial k_i}{\partial z_0}$ \\
\hline
$k_1$ 
& $2(1+e)^2\!\left[r_0 + a(-1+e)\cos d\Phi\right]$ 
& $-2a(-1+e)(1+e)^2 r_0 \sin d\Phi$ 
& $2(1+e)^2 z_0$ \\

$k_2$ 
& $4a(-1+e)(1+e)^2 \sin d\Phi$ 
& $4a(-1+e)(1+e)^2 r_0 \cos d\Phi$ 
& $0$ \\

$k_3$ 
& $-4(1-e^2)\!\left[r_0 + a e \cos d\Phi\right]$ 
& $4ae(1-e^2) r_0 \sin d\Phi$ 
& $-4(1-e^2) z_0$ \\

$k_4$ 
& $-4a(-1+e)^2(1+e)\sin d\Phi$ 
& $-4a(-1+e)^2(1+e) r_0 \cos d\Phi$ 
& $0$ \\

$k_5$ 
& $2(-1+e)^2\!\left[r_0 + a(1+e)\cos d\Phi\right]$ 
& $-2a(-1+e)^2(1+e) r_0 \sin d\Phi$ 
& $2(-1+e)^2 z_0$ \\
\hline
\end{tabular}
\caption{
Derivatives of $k_i$ with respect to $r_0$, $\theta_0$, and $z_0$.}
\label{tab:ki-derivatives}
\end{table}

Hence the three integrals for the accelerations are

\begin{equation}
\begin{aligned}
a_r(t)
&= \int
\frac{
  2 r_0\!\left[\,e(t^4 - 1) - (1 + t^2)^2\,\right]
}{
  \bigl(k_1 + k_2 t + k_3 t^2 + k_4 t^3 + k_5 t^4\bigr)^{3/2}
}\,\mathrm{d}t
\\[6pt]
&\qquad
+ \int
\frac{
  2 a (e^2 - 1)
  \Big[
      (t^4 - 1)\cos(\mathrm{d}\Phi)
    - 2(t + t^3)\sin(\mathrm{d}\Phi)
  \Big]
}{
  \bigl(k_1 + k_2 t + k_3 t^2 + k_4 t^3 + k_5 t^4\bigr)^{3/2}
}\,\mathrm{d}t,
\\[1.2em]
a_{\theta}(t)
&= -\int
\frac{
  2 a (e^2 - 1)(1 + t^2)
  \Big[
      2t\cos(\mathrm{d}\Phi)
    + (t^2 - 1)\sin(\mathrm{d}\Phi)
  \Big]
}{
  \bigl(k_1 + k_2 t + k_3 t^2 + k_4 t^3 + k_5 t^4\bigr)^{3/2}
}\,\mathrm{d}t,
\\[1.2em]
a_z(t)
&= \int
\frac{
  2(1 + t^2)\,[\,-(1 + e) + (e - 1)t^2\,]\,z_0
}{
  \bigl(k_1 + k_2 t + k_3 t^2 + k_4 t^3 + k_5 t^4\bigr)^{3/2}
}\,\mathrm{d}t.
\end{aligned}
\end{equation}

In general, these expressions can be written in terms of sums of the elliptic $\boldsymbol{F},\boldsymbol{E},\boldsymbol{\Pi}$ functions so typically we would have to derive the corresponding continuations as preformed in the previous sections.  Fortunately, in our limit of a full ring all the dependence on the $\boldsymbol{\Pi}$ vanishes, leaving only $\boldsymbol{F},\boldsymbol{E}$ which have simple continuations. We find any acceleration $a_p$, can be written as a linear combination of $\boldsymbol{F} and \boldsymbol{E}$ jumps as

\begin{equation}
a_p \;=\;
\frac{GM(1 - e^2)^{3/2}}{2\pi}\,
\Big[
  c_{F,p}\,\Delta_F
  + c_{E,p}\,\Delta_E
\Big].
\end{equation}

With constants $c_{F,p}$ and $c_{E,p}$ defined as

\begin{subequations}
\begin{align}
c_{F,r}
&=
-\frac{4}{\sqrt{k_5}\,R_2\,\eta_1}
\Big[
    2k_5R_1\!\Big(
        r_0\kappa_1
        + a(e^{2}-1)\big(\eta_3\cos(d\phi) - \eta_2\sin(d\phi)\big)
    \Big)
\\[-2pt]&\qquad\qquad
    -\Big(
        r_0\kappa_2
        - a(e^{2}-1)\big(\eta_5\cos(d\phi) - \eta_4\sin(d\phi)\big)
    \Big)
\Big],
\\[1.0em]
c_{E,r}
&=
\frac{8\sqrt{k_5}R_2}{\eta_1}
\Big[
    r_0\kappa_1
    + a(e^{2}-1)\eta_3\cos(d\phi)
    - a(e^{2}-1)\eta_2\sin(d\phi)
\Big],
\\[1.0em]
c_{F,\theta}
&=
\frac{4a(e^{2}-1)}{\sqrt{k_5}\,R_2\,\eta_1}
\Big[
    \eta_4\cos(d\phi) + \eta_5\sin(d\phi)
\\[-2pt]&\qquad\qquad
    + 2k_5R_1\big(-\eta_2\cos(d\phi) - \eta_3\sin(d\phi)\big)
\Big],
\\[1.0em]
c_{E,\theta}
&=
-\frac{8a\sqrt{k_5}(e^{2}-1)R_2}{\eta_1}
\Big[
    \eta_2\cos(d\phi) + \eta_3\sin(d\phi)
\Big],
\\[1.0em]
c_{F,z}
&=
\frac{4z_0}{R_2\,\eta_1}
\Big[
    2k_5R_1\kappa_1 + \kappa_2
\Big],
\\[1.0em]
c_{E,z}
&=
\frac{8\sqrt{k_5}R_2z_0 \kappa_1}{\eta_1}
\end{align}
\end{subequations}

Finally the remaining free variables can be written in terms of the $r_i$ and $k_i$ as follows. The $R_i$ are real combinations of the roots. 

\begin{subequations}
\begin{align}
R_1 &= r_2 r_3 + r_1 r_4, \\
R_2 &= \sqrt{(r_1 - r_3)(r_2 - r_4)}
\end{align}    
\end{subequations}

\begingroup
\begin{align}
\tag{\theequation}\refstepcounter{equation}\label{eq:kappa}%
\kappa_1 &= 3(-1+e)k_2^3k_4 + (1+e)k_3^2(k_4^2-4k_3k_5) \\[-0.25em]\notag
&\quad + 2k_1^2\!\Big(9(-1+e)k_4^2 + 8(k_3 - e k_3 - 4k_5)k_5\Big) \\[-0.25em]\notag
&\quad + 2k_1\!\Big(2(-1+e)k_3^3 - 3k_3k_4^2 + 8k_3^2k_5
- 3(1+e)k_4^2k_5 + 8(1+e)k_3k_5^2\Big) \\[-0.25em]\notag
&\quad + k_2^2\!\Big(-(-1+e)k_3^2 + 2k_4^2 - 6k_3k_5
- 6k_5(k_1 - e k_1 + 3(1+e)k_5)\Big) \\[-0.25em]\notag
&\quad + k_2k_4\!\Big(-14(-1+e)k_1k_3 + 8k_1k_5
- (1+e)(3k_4^2 - 14k_3k_5)\Big), \\[0.4em]\notag
\kappa_2 &= k_2^3k_4((-1+e)k_3 - 6k_5) - 9(-1+e)k_2^4k_5 \\[-0.25em]\notag
&\quad + k_2^2\!\Big((k_1 - e k_1 - 2k_3)k_4^2
+ (44(-1+e)k_1k_3 + 8k_3^2 + (1+e)k_4^2)k_5 \\[-0.25em]\notag
&\qquad - 12(4k_1 + k_3 + e k_3)k_5^2\Big)
+ k_2k_4\!\Big(-64(-1+e)k_1^2k_5
+ (1+e)k_3(-k_4^2 + 4k_3k_5) \\[-0.25em]\notag
&\qquad + k_1(-4(-1+e)k_3^2 - 6k_4^2 + 48k_3k_5 + 64(1+e)k_5^2)\Big) \\[-0.25em]\notag
&\quad + k_1\!\Big(128(-1+e)k_1^2k_5^2
+ 4k_1(3(-1+e)k_3k_4^2
- 4(2(-1+e)k_3^2 + 3k_4^2)k_5 \\[-0.25em]\notag
&\qquad + 32k_3k_5^2 - 32(1+e)k_5^3)
+ (k_4^2 - 4k_3k_5)\big(9(1+e)k_4^2
+ 8k_3(k_3 - (1+e)k_5)\big)\Big). \notag
\end{align}
\endgroup

\begingroup
\begin{align}
\tag{\theequation}\refstepcounter{equation}\label{eq:etas}%
\eta_1 &= 27k_2^4 k_5^2
+ 2k_2^3(2k_4^3 - 9k_3k_4k_5)
+ 2k_1k_2k_4(-9k_3k_4^2 + 40k_3^2k_5 + 96k_1k_5^2) \\[0.1em]\notag
&\quad + k_2^2(-k_3^2k_4^2 + 4k_3^3k_5 + 6k_1k_4^2k_5 - 144k_1k_3k_5^2) \\[0.1em]\notag
&\quad + k_1\!\Big(4k_3^3k_4^2 - 16k_3^4k_5 - 144k_1k_3k_4^2k_5
+ 128k_1k_3^2k_5^2 + k_1(27k_4^4 - 256k_1k_5^3)\Big), \\[0.4em]\notag
\eta_2 &= -9k_2^3k_5 + k_2^2k_4(k_3 + 3k_5)
+ k_2k_3(k_4^2 - 4k_3k_5) \\[0.1em]\notag
&\quad + k_1k_2(3k_4^2 + 16(2k_3 - 3k_5)k_5)
- k_1k_4(4k_3^2 + 9k_4^2 + 48k_1k_5 - 32k_3k_5), \\[0.4em]\notag
\eta_3 &= 3k_2^3k_4 + 2k_1^2(9k_4^2 - 8k_3k_5)
+ k_3^2(k_4^2 - 4k_3k_5) \\[0.1em]\notag
&\quad + k_2(-14k_1k_3k_4 - 3k_4^3 + 14k_3k_4k_5)
+ 2k_1(2k_3^3 - 3k_4^2k_5 + 8k_3k_5^2) \\[0.1em]\notag
&\quad - k_2^2(k_3^2 + 6k_5(-k_1 + 3k_5)), \\[0.4em]\notag
\eta_4 &= 4\Big[
k_4^2(k_2^3 - 4k_1k_2k_3 + k_2^2k_4 + 3k_1(3k_1 - k_3)k_4) \\[-0.25em]\notag
&\quad + (-3k_2^3k_3 + k_2^2(k_1 - 4k_3)k_4
+ 4k_1k_3(-8k_1 + 3k_3)k_4 + k_1k_2(12k_3^2 + k_4^2))k_5 \\[-0.25em]\notag
&\quad + (9k_2^3 - 32k_1k_2k_3 + 16k_1^2(k_2 + k_4))k_5^2
\Big], \\[0.4em]\notag
\eta_5 &= k_4\!\Big(-k_2^3k_3 + k_1k_2^2k_4 + k_2k_3(4k_1k_3 + k_4^2)
- 3k_1k_4(4k_1k_3 + 3k_4^2)\Big) \\[-0.25em]\notag
&\quad + \Big(9k_2^4 + k_2(64k_1^2k_4 - 4k_3^2k_4)
- k_2^2(44k_1k_3 + k_4^2)
+ 4k_1k_3(8k_1k_3 + 11k_4^2)\Big)k_5 \\[-0.25em]\notag
&\quad - 4(-3k_2^2k_3 + 8k_1(4k_1^2 + k_3^2 + 2k_2k_4))k_5^2
+ 128k_1^2k_5^3. \notag
\end{align}
\endgroup

\begin{figure}[!ht]
\centering
    \includegraphics[width=\linewidth, keepaspectratio]{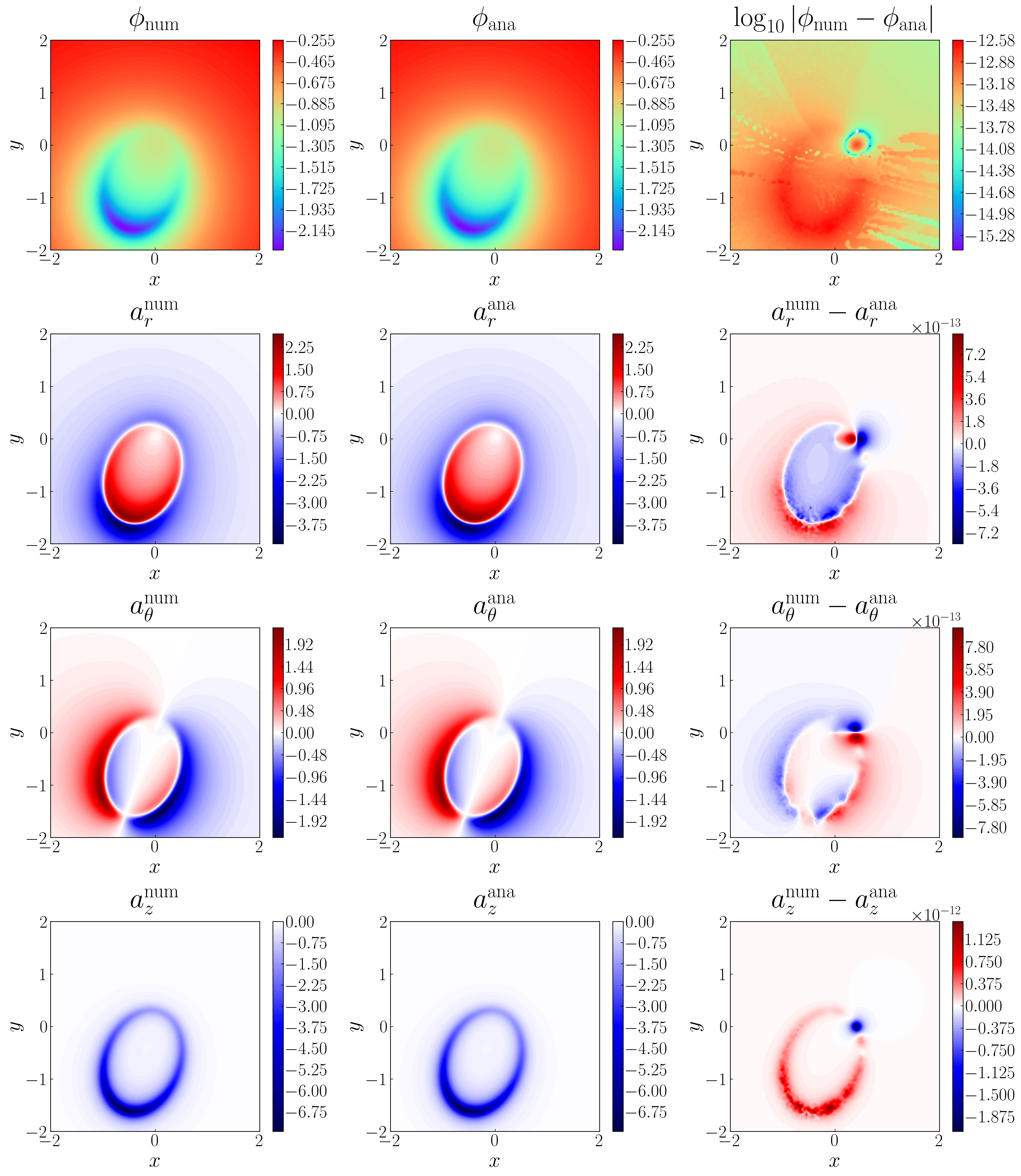}
    \caption{The top left panel shows the result of a numerical integration $\phi_{num}$, the top center panel shows the analytic form $\phi_{ana}$  and the top right panel shows the log of their difference. The lower rows shows the accelerations in $r$, $\theta$,$z$ directions.   All quantities were computed assuming $G = M = a = 1$, $e=0.7$, $\omega=70^\circ$ and $z_0=0.1$ }
    \label{fig:eccringpl}
\end{figure}

While complex, these expressions have a few desirable features. The first is that the only variables involving the roots $r_i$ are $m$, $R_1$ and $R_2$.  All of these are symmetric combinations of roots and are real.  The $\eta_i$ and $\kappa_i$ are all simply polynomial functions of $k_i$ which are themselves directly related to the orbital parameters. Finally, the two jumps $\Delta_F$ and $\Delta_E$ are both independent of where the jump occurs, hence no $t_i$ need to be computed and no patches applied.  In contrast to most of the expressions considered thus far, both jumps are simply complete elliptic integrals of a single real parameter, this simplicity opens up a wide variety of methods for rapid evaluation including rational or polynomial series, iterative methods or interpolation over a grid.  

The potential and accelerations can be seen in Fig \ref{fig:eccringpl}.  We can see that, in contrast to the case of the uniform ellipse, the angle of greatest radial acceleration (both outwards and inwards) is aligned in the case of the time-averaged ellipse.  The $a_z$ acceleration is now asymmetric, reflecting the fact that an orbiting body spends most of it's time at apoapsis. Finally, the sign of $a_\theta$ accelerations are now not totally dependent on the $\omega$, but on the precise location within the ring. While near apoapsis it retains the same structure as the uniform ellipse, it flips sign near periapsis.  Now that a closed form is known, these and other properties of the solution can be explored further algebraically.

\section{Data Availability}
The expressions discussed in this paper can be complex and difficult to re-implement for the purposes of reproduction.  To ease verification of our results we provide code to reproduce the figures in this paper, this code includes a python implementation of our expressions.   A repository for this code can be found here at \url{https://github.com/r-zachary-murray/analyticpotentials}.

\section{Acknowledgments}

This work was supported by the French government through the France 2030 investment plan managed by the National Research Agency (ANR), as part of the Initiative of Excellence of Université Côte d’Azur under reference number ANR-15-IDEX-01.

\begin{appendices}

\section{Deriving Elliptic Jumps}\label{secA1}

In this section we derive the jumps in the $\boldsymbol{\Pi}$ function. Since there is a simultaneous branch cut and flip of sign in the case of $\Pi_1$, the left and right sided limits must be added together $\Delta_{\Pi_1}  = \lim_{t \to t_{3}^{-}} \boldsymbol{\Pi}_{1}(t) \;+\; \lim_{t \to t_{3}^{+}} \boldsymbol{\Pi}_{1}(t)$.  We can derive the appropriate expressions by it by noting the $\boldsymbol{\Pi}$ function can be represented in terms of Carlson functions through

\begin{equation}
\begin{aligned}
\boldsymbol{\Pi}(n;\,\phi \mid m)
&= \sin\phi\;
   \boldsymbol{R_F}\!\bigl(\cos^{2}\!\phi,\; 1 - m\sin^{2}\!\phi,\; 1\bigr)
\\[0.4em]
&\quad +\;
   \frac{n\,\sin^{3}\!\phi}{3}\;
   \boldsymbol{R_J}\!\bigl(\cos^{2}\!\phi,\; 1 - m\sin^{2}\!\phi,\; 1,\; 1 - n\sin^{2}\!\phi\bigr).
\end{aligned}
\end{equation}

We consider both jumps separately.  The jump in the $R_F$ component is simple, and is simply twice the value at that point $-2\Re{\boldsymbol{R_f}(1-s^2,1-m s^2,1)}$.  The jump of $\boldsymbol{R_J}$ is somewhat more complicated.  If we define as shorthand 
\begin{equation}
x=\cos^2\phi,\quad
p=1-m\sin^2\phi,\quad
y_\pm=1-n\sin^2\phi,
\end{equation}

Then

\begin{equation}
\begin{aligned}
\boldsymbol{\Pi}(n;\,\phi \mid m)
&= \sin\phi\;
   \boldsymbol{R_F}\!\bigl(\cos^{2}\!\phi,\; 1 - m\sin^{2}\!\phi,\; 1\bigr)
\\[0.4em]
&\quad +\;
   \frac{n\,\sin^{3}\!\phi}{3}\;
   \boldsymbol{R_J}\!\bigl(\cos^{2}\!\phi,\; 1 - m\sin^{2}\!\phi,\; 1,\; 1 - n\sin^{2}\!\phi\bigr).
\end{aligned}
\end{equation}

with

\begin{equation}
\beta=(n-1)\sin^2\phi,\qquad
\gamma=n\sin^2\phi,\qquad
\alpha=(n-m)\sin^2\phi.
\end{equation}

This is equal to, by definition

\begin{align}
\boldsymbol{R_J}(\beta,0,\gamma,\alpha)
&= \frac{3}{2}\,\!\int_{0}^{\infty}
\frac{du}{u\,\sqrt{(u+\beta)(u+\alpha)(u+\gamma)}}
\end{align}

Putting everything together, we find the jump for both $\Pi$ functions is simply 

$\Delta_{B,\Pi} = - \frac{2 n s^3}{3} R_J(\beta,0,\gamma,\alpha) + 2 \Re{\boldsymbol{R_f}(1-s^2,1-m s^2,1)}$.

Finally we have the case of the pole jump in the  $\boldsymbol{\Pi}$ function. Once again we split the $\boldsymbol{\Pi_2}$ function into its Carlson components, and note that $\boldsymbol{R_F}$ has no jump at $t=t_2$, therefore the entire jump must come from $\boldsymbol{R_J}$

\begin{align*}
\boldsymbol{R_J}(x,y,1,p)
&= \frac{3}{2}\int_{0}^{\infty} \frac{dt}{(t+p)\sqrt{(t+x)(t+y)(t+1)}}.
\end{align*}

When $p=-a<0$, the integrand has a simple pole at $t=a$ on the path of integration.  As typical, we can deform the in a small loop about the pole to write the jump as

\begin{equation}
\begin{aligned}
\boldsymbol{R_J}^{(-)} - \boldsymbol{R_J}^{(+)}
&= \oint_{|t-a|=\varepsilon}
   \frac{\tfrac{3}{2}}{(t-a)\sqrt{(t+x)(t+y)(t+1)}}\,dt
\\[0.6em]
&=\; (2\pi i)\,\frac{3}{2}\,
   \frac{1}{\sqrt{(t+x)(t+y)(t+1)}}.
\end{aligned}
\end{equation}

The jump is therefore: 
\begin{equation}
    \Delta_{P,\Pi} = \boldsymbol{R_J}^{(-)} - \boldsymbol{R_J}^{(+)} = \frac{3\pi i}{\sqrt{(a+x)(a+y)(a+1)}}.
\end{equation}

\end{appendices}

\bibliography{sn-bibliography}

\end{document}